\documentclass[12pt]{article}

\usepackage{amssymb}
\usepackage{pstricks}

\newcommand{\nc}{\newcommand}
\nc{\CC}{\mathbb C}
\nc{\NN}{\mathbb N}
\nc{\RR}{\mathbb R}
\nc{\ZZ}{\mathbb Z}
\nc{\mcH}{\mathcal{H}}
\nc{\mfH}{\mathfrak{H}}
\nc{\mcU}{\mathcal{U}}
\nc{\dom}{\mbox{dom}}
\nc{\Ide}{\mbox{Id}_E}
\nc{\ess}{\mbox{ess}}
\nc{\supp}{\mbox{supp}}
\nc{\dist}{\mbox{dist}}
\nc{\Hpos}{\mcH_{\mbox{\scriptsize pos}}}
\nc{\Hneg}{\mcH_{\mbox{\scriptsize neg}}}
\nc{\Hposneg}{\mcH_{\mathop{\mbox{\scriptsize pos}}\limits_{\mathop{\mbox{\scriptsize neg}}}}}

\renewcommand{\parallel}{\|}

\begin{document}

\title{Localization and Semibounded Energy - \\
A Weak Unique Continuation Theorem}

\author{Christian B\"ar}

\date{18. June 1999}
\maketitle

\begin{abstract}
\noindent
Let $D$ be a self-adjoint differential operator of Dirac type
acting on sections in a vector bundle over a closed Riemannian manifold $M$.
Let $\mcH$ be a closed $D$-invariant subspace of the Hilbert space of square
integrable sections.
Suppose $D$ restricted to $\mcH$ is semibounded.
We show that every element $\psi\in\mcH$ has the weak unique
continuation property, i.e. if $\psi$ vanishes on a nonempty open subset of
$M$, then it vanishes on all of $M$.


\end{abstract}

\section{Introduction}
In relativistic quantum mechanics an electron at a fixed time $t=0$ is described by
a wave function (a spinor) $\psi_0 : \RR^3\rightarrow\CC^4$ normalized by
$\parallel \psi_0\parallel_{L^2(\RR^3)}=1$.
Usually one interprets $|\psi_0(x)|^2$ as the probability density to find the
electron at the point $x$ at time $t=0$.
The dynamics are given by
$$
\psi(t,x) = \left(e^{itD}\psi_0\right)(x)
$$
where $D$ is the spacial Dirac operator (possibly coupled to an external
field).
The spectrum of $D$ is unbounded to the left and to the right which causes some
interpretational difficulties: 
``But an interacting particle may exchange energy with its environment, and 
there would then be nothing to stop it cascading down to infinite
negative energy states, emitting an infinite amount of energy in the process''
\cite[p.29]{ryder96a}.
Of course, this is not a realistic scenario.

The problem is usually overcome by splitting $L^2(\RR^3,\CC^4)$ into the 
spectral subspaces of positive and negative energy
\begin{equation}
L^2(\RR^3,\CC^4)= \Hpos \oplus \Hneg
\label{spalt}
\end{equation}
where $\Hposneg$ is the subspace corresponding to the positive/negative part 
of the spectrum of $D$.
Here we assume for simplicity that $0$ is not in the spectrum.
Now one requires a wave function of the electron to lie in $\Hpos$.
A $\psi_0\in\Hneg$ would be interpreted as a wave function for the antiparticle, the positron.
For the free Dirac operator (without external field) one can show 
\cite[Cor.17]{thaller92a} that any $\psi_0\in\Hpos$ (or $\Hneg$) has the 
{\it weak unique continuation property}, i.e. if $\Omega\subset\RR^3$ is 
nonempty and open, then
$$
\psi_0|_{\Omega}=0 \quad \Longrightarrow \quad \psi_0 = 0 \quad \mbox{on } \RR^3.
$$
This means in particular, that a free electron can never be localized, i.e. the support
of $\psi_0$ cannot be contained in a compact set.
The proof given in \cite[Cor.17]{thaller92a} relies on the explicit form of the free Dirac operator
on $\RR^3$ and its Fourier transform.
We will see that the weak unique continuation property of elements of 
semibounded spectral subspaces is a general fact for operators of Dirac type 
(see next section for a definition) and for even more
general operators at least if the underlying manifold is closed.
Here ``closed'' means compact, connected, and without boundary.

{\bf Theorem.}
{\it
Let $M$ be a closed Riemannian manifold, let $E\rightarrow M$ be a Hermitian
vector bundle and let $D$ be a self-adjoint differential operator of Dirac
type acting on sections of $E$.
\newline
Let $\mcH\subset L^2(M,E)$ be a closed subspace, such that $D(\mcH\cap \dom(D))
\subset\mcH$ and $D|_{\mcH\cap dom(D)}$ is self-adjoint in $\mcH$.
Suppose that the restriction of $D$ to $\mcH$ is semibounded.

Then if $\varphi\in\mcH$ vanishes on a nonempty open subset $\Omega\subset M$
it actually vanishes on all of $M$.}

In particular, if we choose $\mcH$ to be an eigenspace, then this says that eigensections of $D$ have
the weak unique continuation property.
This is nontrivial but well-known, see e.g.\
\cite{booss99ppa,booss-wojciechowski93a}, and we will in fact use this
special case in our proof.

It should be mentioned that a splitting as in (\ref{spalt}) also occurs in
purely mathematical context.
In order to make the Dirac operator on a compact manifold with boundary 
Fredholm one imposes the famous Atiyah-Patodi-Singer boundary conditions
\cite{atiyah-patodi-singer75a}.
These conditions simply mean that the restriction of the spinor to the 
boundary must lie in $\mcH = \Hneg$.

Let us emphasize the difference of our theorem to the standard results on the 
weak unique continuation property.
Usually, one requires $\varphi$ to satisfy a differential equation or at least 
a differential inequality of the kind
\begin{equation}
|\Delta\varphi|\le C_1 \cdot |\nabla\varphi| + C_2 \cdot |\varphi|
\label{dug}
\end{equation}
or variations thereof \cite{aronszajn57a,kazdan88a}. 
Here $\Delta$ is an elliptic second-order differential operator with scalar 
principal symbol. 
For $\Delta= D^2$ this shows in particular, that the theorem is true for 
eigensections $\varphi$ of $D$.
In contrast, in our theorem $\varphi$ does not satisfy a differential 
inequality.
The assumption of being in $\mcH$ could rather be called a {\it spectral inequality} on $\varphi$.
In contrast to a differential inequality this is no longer a local condition.

{\bf Acknowledgements.}
The idea to this note arose from discussions in a seminar jointly organized
by mathematicians and phycisists.
It is a particular pleasure to thank 
H.\ R\"omer for helpful hints and valuable insight.

\section{Some Preparations}

Let $M$ be a closed Riemannian manifold, let $E\rightarrow M$ be a Hermitian
vector bundle over $M$.
Denote the Hermitian metric by $\langle \cdot , \cdot \rangle$.
Let $D:C^{\infty}(M,E)\rightarrow C^{\infty}(M,E)$ be a formally self-adjoint differential operator
of first order.
We call $D$ of {\it Dirac type} if its principal symbol $\sigma_D$ satisfies
the Clifford relations, i.e.
$$
\sigma_D (\xi)\circ\sigma_D(\eta)+\sigma_D(\eta)\circ\sigma_D(\xi)=2g(\xi,\eta)\cdot\Ide
$$
for all $\xi$, $\eta\in T^*_p M$, $p\in M$.
Then $D$ is an {elliptic} differential operator, essentially self-adjoint on
$C^{\infty}(M,E)$ in $L^2(M,E)$.
For example, a generalized Dirac operator in the sense of Gromov and Lawson 
\cite{gromov-lawson83a} is of Dirac type.

Let $\mcH\subset L^2(M, E)$ be a closed subspace, invariant under $D$, i.e.
$D(\mcH\cap\dom(D))\subset\mcH$, $\mcH\cap\dom(D)$ is dense in $\mcH$ and
$D|_{\mcH \cap dom(D)} =: D|_{\mcH}$ is self-adjoint.
Let $\{\lambda_j\}$ be the spectrum of $D|_{\mcH}$ and let $\{\varphi_j\}$ be
the corresponding eigensections, normalized by $\parallel \varphi_j
\parallel_{L^2 (M,E)}=1$.

We define sections $\varphi^*_j$ in the dual bundle $E^*$ by
$$
\varphi^*_j (x) (\psi) := \langle \varphi_j(x), \psi\rangle
$$
for all $\psi\in E_x$.
Then the integral kernel of the operator $e^{i z D|_{\mcH}}$ is defined by
$$
q_z (x,y) := \sum\limits_j e^{i z \lambda_j} \varphi_j(x)\otimes\varphi^*_j(y),
$$
$z\in\CC$, $x,y\in M$.
By $H^k$ we denote the Sobolev space of $L^2$-sections whose derivatives up to
order $k$ are again $L^2$.
For each $z$ we consider $q_z$ as a section in the exterior tensor product
$E \boxtimes E^*\rightarrow M\times M$ where $(E\boxtimes E^*)_{(x,y)} = E_x \otimes E^*_y$.

{\bf Lemma 1.}
{\em If $D|_{\mcH}$ is bounded from below, then the series $q_z$ converges
absolutely and locally uniformly for $z\in\{\zeta\in\CC\mid\Im(\zeta)>0\} =:
\mfH$ in each Sobolev space $H^k(M\times M, E\boxtimes E^*)$.}

{\em Proof.}
If $D|_{\mcH}$ is bounded from below, then only finitely many eigenvalues $\lambda_j$ are nonpositive.
Hence we may assume $0<\lambda_1\le\lambda_2\le\lambda_3\le\ldots$.
By ellipticity of $D$ there is a constant $C_1>0$ s.t.
\begin{eqnarray*}
\parallel \varphi_j\parallel_{H^k (M,E)}
&\le&
C_1 \cdot
\left\{
\parallel \varphi_j\parallel_{L^2 (M,E)} +
\parallel D^k \varphi_j\parallel_{L^2 (M,E)}
\right\} \\
&=& C_1 \cdot \left(1+\lambda^k_j\right).
\end{eqnarray*}
Let $\Im (z) \ge \epsilon >0$.
Then
\begin{eqnarray*}
\parallel q_z\parallel_{H^k(M\times M, E\boxtimes E^*)}
&\le&
\sum\limits_j e^{-\Im (z)\lambda_j}
\cdot \parallel\varphi_j\parallel_{H^k(M, E)}
\cdot \parallel\varphi_j^*\parallel_{H^k(M, E^*)}\\
&\le&
C^2_1 \cdot\sum\limits_j e^{-\epsilon\lambda_j}\cdot\left(1+\lambda^k_j\right)^2 \\
&\le&
C_2\cdot\sum\limits_j e^{-\epsilon\lambda_j/2}
\end{eqnarray*}
since the function $\lambda \mapsto e^{-\epsilon\lambda/2} \cdot \left(1+
\lambda^k\right)^2$ is bounded for $\lambda\in(0, \infty)$.
From Weyl's asymptotic formula \cite[Cor. 2.43]{berline-getzler-vergne91a} we 
know
$$
\lambda_j\ge C_3 \cdot j^{\alpha}
$$
for some $\alpha>0$.
Note that the eigenvalues of $D|_\mcH$ grow at least as fast as those of $D$.
Hence
$$
\parallel q_z\parallel_{H^k(M\times M, E\boxtimes E^*)} \le C_2 \cdot
\sum\limits_j e^{-C_4\cdot j^{\alpha}} < \infty
$$
$\hfill\square$

{\bf Corollary.}
{\em If $D|_{\mcH}$ is bounded from below, then
\begin{eqnarray*}
\mfH &\rightarrow& H^k (M\times M, E\boxtimes E^*), \\
z&\mapsto& q_z,
\end{eqnarray*}
is holomorphic for each $k\in\NN$ and $q_z(x,y)$ is smooth in
$(z,x,y) \in \mfH \times M\times M$.}
$\hfill\square$

Next we need a technical uniqueness lemma for holomorphic functions.

{\bf Lemma 2.}
{\em Let $f: \overline{\mfH} = \mfH\cup\RR\rightarrow\CC$ be a continuous
function and let its restriction $f|\mfH$ be holomorphic.
If there is a nonempty open interval $I\subset \RR$ such that $f|_I = 0$,
then $f$ vanishes on all of $\overline{\mfH}$.}

{\em Proof.}
Pick $t$ in the interior of $I$ and a small disk $\Delta \subset \CC$ with
center $t$ such that $\Delta\cap\RR\subset I$.


\begin{center}
\pspicture(1,0)(14,6)

\pspolygon[linewidth=1pt,linecolor=lightgray,fillstyle=solid,fillcolor=lightgray](2,3)(12,3)(12,6)(2,6)
\psline[linewidth=1pt](2,3)(12,3)
\psline[linewidth=2pt](4.5,3)(9.5,3)
\pscircle[linewidth=1pt](7,3){1.5}

\rput(3,5){\psframebox*[framearc=0.5]{$\mathfrak{H}$}}
\rput(6.9,3.9){\psframebox*[framearc=0.5]{$\Delta$}}
\rput(9.2,3.5){\psframebox*[framearc=0.5]{$I$}}
\rput(12.5,3){{$\mathbb{R}$}}

\psdots[dotsize=5pt,dotstyle=*](7,3)
\rput(7,2.7){{$t$}}

\endpspicture
\vspace{-1cm}
\end{center}
\begin{center}
\bf Fig.~1
\end{center}


By Schwarz's reflection principle we can extend $f$ holomorphically to
$\Delta$.
Since $f$ vanishes on $\Delta\cap\RR$ it must vanish on all of $\Delta$ and therefore on all of
$\mathfrak{H}$. $\hfill\square$

We need one last tool known as {\it finite propagation speed.}

{\bf Lemma 3.}
{\em Let $D:C^{\infty} (M,E)\rightarrow C^{\infty}(M,E)$ be a self-adjoint
differential operator of Dirac type, let $\psi\in L^2(M,E)$.
Then for all $t\in\RR$
$$
\ess - \supp \left(e^{it D}\psi\right)\subset\mcU_{|t|}(\ess - \supp(\psi))
$$
where $\mcU_r(A)=\{x\in M\ |\ \dist(x,A)\le r\}$ is the $r$-neighborhood of
the subset $A\subset M$.}

The lemma says that the support of $\psi$ grows at most with speed one.
See e.g. \cite[Prop. 5.5]{roe88a} for a proof.

\section{Proof of the Theorem}

Now we are able to prove the theorem.
Replacing $D$ by $-D$ if necessary we may w.l.o.g.\ assume that $D|_\mcH$
is bounded from below.
Let $\psi\in\mathcal{H}$, $\Omega\subset M$ open, $\Omega\not=\emptyset$, and
$\psi|_{\Omega}=0$.
We want to show that $\psi=0$.
\newline
Let $P_{\Omega}$ be the projection in $L^2(M, E)$ defined by restriction to
$\Omega$,
$$
\left(P_{\Omega}\varphi\right)(x):=
\left\{
\begin{array}{cc}
\varphi(x),& x\in\Omega\\
0,& x\in M-\Omega.
\end{array}
\right.
$$
Pick any nonempty open subset $\Omega' \subset\subset\Omega$.
By Lemma 3 there is an $\epsilon>0$, such that
$$
e^{i t D} \psi|_{\Omega'} = 0
$$
for all $t\in[0, \epsilon)$.
Fix $\varphi\in L^2(M,E)$ and define
\begin{eqnarray*}
f_{\varphi}(z) &:=& \left(\varphi,P_{\Omega'} e^{i z D}\psi\right)_{L^2(M,E)}\\
&=& \left(P_{\Omega'} \varphi,  e^{i z D}\psi\right)_{L^2(M,E)}\\
&=& \left(P_{\Omega'} \varphi,  e^{i z D|_{\mcH}}\psi\right)_{L^2(M,E)}.
\end{eqnarray*}
By the corollary to Lemma 1 $f_\varphi$ is holomorphic on $\mfH$.
Since $D|_\mcH$ is bounded from below, the functions $g_z(\lambda) =
e^{iz\lambda}$ are uniformly bounded on the spectrum of $D|_\mcH$ for all
$z\in \overline{\mfH}$.
Moreover, for $z_j \to z$ we have $g_{z_j} \to g_z$ locally uniformly.
Thus $\mbox{s}-\lim_j g_{z_j}(D|_\mcH) = g_z(D|_\mcH)$.
Therefore $f_{\varphi}$ is continuous on $\overline{\mfH}$.

Since $f_{\varphi}$ vanishes on $[0, \epsilon)$ Lemma 2 implies $f_{\varphi}=0$ on
$\overline{\mfH}$.
Since $\varphi$ is arbitrary this shows
$$
P_{\Omega'}  e^{i z D}\psi = 0
$$
for all $z\in\overline{\mfH}$.
In particular, for $z=it$, $t>0$, this means
$$
P_{\Omega'}  e^{-t D}\psi = 0.
$$
It follows that $P_{\Omega'}  e^{-t(D-\lambda_1)}\psi = e^{t\lambda_1}
P_{\Omega'}  e^{-t D}\psi = 0$ for all $t>0$.
Let $P_{\lambda_1}$ be the projection in $L^2(M,E)$ onto the $\lambda_1$-eigenspace for $D$.
Then
$$
0=\lim\limits_{t\rightarrow\infty} P_{\Omega'} e^{-t(D-\lambda_1)}\psi =
P_{\Omega'} \lim\limits_{t\rightarrow\infty} e^{-t(D-\lambda_1)}\psi
=
P_{\Omega'} P_{\lambda_1}\psi.
$$
As an eigensection of $D$, $ P_{\lambda_1}\psi$ has the weak unique continuation property, hence
$P_{\Omega'} P_{\lambda_1}\psi=0$ implies
$$
P_{\lambda_1}\psi=0.
$$
Now we can replace $\lambda_1$ by $\lambda_2$ and repeat the argument to obtain
$$
P_{\lambda_2}\psi=0
$$
and inductively
$$
\psi=0.
$$
$\hfill\square$

\section{Concluding Remarks}

The assumption that the operator $D$ is of Dirac type was made mostly for 
convenience.
In fact, it was used in a rather inessential way. 
Lemma 1 holds for any self-adjoint elliptic differential operator defined 
over a closed manifold while in Lemma 3 even ellipticity could be dispensed 
with.
In the proof of the theorem itself we used the unique continuation property 
of eigensections of Dirac type operators.
Summing up we see that 

{\it the theorem holds for all self-adjoint elliptic differential operators 
of first order defined over a closed manifold whose eigensections are known 
to have the weak unique continuation property}.

Note that by (\ref{dug}) this is automatic if $D^2$ has scalar principal 
symbol. 
But that is equivalent to $D$ being of Dirac type for some Riemannian metric.

One may also try to relax the condition that the underlying manifold is closed.
In fact, the manifold for which the problem was originally considered, namely $\RR^3$, is not closed.
Therefore we would like to replace ``closed'' by ``complete''.
Closedness of $M$ has been used in Lemma 1 since it guarantees discreteness of the spectrum and Weyl's
asymptotic law.
Discreteness of the spectrum is also important for the induction in the proof of the theorem.
Whether or not the theorem also holds for complete manifolds has to be seen.
In case of $\RR^3$ this would imply that even in an external field the electron has no localized states.

For eigensections of a Dirac operator much more is known than just the weak unique continuation property.
Namely, if $\varphi$ satisfies $D\varphi=\lambda\varphi$, then the zero set 
of $\varphi$ has Hausdorff-dimension $\le n-2$ where $n$ is the dimension of 
the manifold \cite{baer97b,baer98ppa}.
We may ask if this is still true for $\varphi$ in our spectral subspace, $\varphi\in\mcH$.
The answer however is no.
Look at the following simple example: 

Let $M=S^1=\RR / 2\pi\ZZ$, let $E$ be the trivial complex line bundle over $M$,
 let $D=i\frac{d}{dt}$, and let $\varphi(t)=e^{-it} + e^{-2it}$.
Then $\varphi$ is the sum of two eigenfunctions, hence lies in a subspace of 
$L^2(S^1, \CC)$ on which $D$ is bounded from below and from above.
But $\varphi$ has a zero at $t=\pi$, thus the codimension of the zero set is $1$ only.



\providecommand{\bysame}{\leavevmode\hbox to3em{\hrulefill}\thinspace}

\vspace{1cm}

\parskip0ex

Mathematisches Institut

Universit\"at Freiburg

Eckerstr.~1

79104 Freiburg

Germany

\vspace{0.5cm}

E-Mail:
{\tt baer@mathematik.uni-freiburg.de}

WWW:
{\tt http://web.mathematik.uni-freiburg.de/home/baer}

\end{document}